\documentclass[12pt,letterpaper]{JHEP3}

\input{epsf}
\usepackage{epsfig}
\usepackage{amsmath}
\usepackage{cite}

\def\pa{\partial}

\def\to{\rightarrow}
\def\be{\begin{equation}}
\def\ee{\end{equation}}
\def\bea{\begin{eqnarray}}
\def\eea{\end{eqnarray}}
\def\nonu{\nonumber \\{}}
\def\half{{1 \over 2}}
\def\ca{{\cal{A}}}
\def\cf{{\cal{F}}}
\def\cg{{\cal{G}}}

\def\cl{{\cal{L}}}
\def\cm{{\cal{M}}}
\def\cn{{\cal{N}}}

\def\sF{{{\rm F}\!\!\!\!\hskip.8pt\hbox{\raise1pt\hbox{/}}\,}}

\def\a{\alpha}
\def\b{\beta}
\def\c{\gamma}

\def\f{\phi}
\def\c{\chi}

\def\m{\mu}
\def\n{\nu}
\def\o{\omega}
\def\p{\pi}
\def\q{\theta}
\def\r{\rho}
\def\s{\sigma}
\def\t{\tau}

\def\x{\xi}

\def\F{\Phi}

\def\O{\Omega}

\title{\begin{center}Near-horizon microstates of the D1-D5-P black hole\end{center}}
\author{\centerline{Joris Raeymaekers$^\ast\;^\dagger$}\\
\centerline{$^\ast$Department of Physics, University of Tokyo,}
\centerline{Hongo 7-3-1, Bunkyo-ku, Tokyo 113-0033, Japan.}
\centerline{$^\dagger$ Institute for Theoretical Physics, K.U.
Leuven,} \centerline{Celestijnenlaan 200D, B-3001 Leuven,
Belgium.}

\centerline{} \centerline{}
\bigskip
\centerline{{\rm
E-mail}:\email{joris@hep-th.phys.s.u-tokyo.ac.jp}}}

\abstract{We construct probe solutions in the attractor background of the five-dimensional D1-D5-P black hole
which represent near-horizon microstates in the limit of large D1-charge.
These generalize the corresponding solutions considered by Gaiotto, Strominger and Yin  for the
4-dimensional D0-D4 black hole. Using U-duality and a 4D-5D connection, we argue that
the relevant configurations are bound states of D1-branes that have expanded through the Myers effect to form a
Kaluza-Klein monopole wrapping the black hole horizon. We show that these branes experience a magnetic
field on their moduli space, and that the degeneracy of lowest Landau levels reproduces the Bekenstein-Hawking
entropy.}


\preprint{ UT-07-34}

\begin{document}
\section{Introduction and summary}
The microscopic accounting of the Bekenstein-Hawking entropy of black holes is one of the  successes
of string theory \cite{Strominger:1996sh,Callan:1996dv}. In most examples, such an accounting starts from the observation
that the entropy
(or supersymmetric index) doesn't depend on a coupling parameter  and subsequently varying  the
coupling to a regime where a perturbative calculation is possible.

In an alternative, and in some sense more direct,
 approach, Gaiotto, Strominger and Yin have proposed to account for the black hole entropy from counting the supersymmetric
 bound states of the constituent D-branes as probes in the near-horizon geometry of the black hole  \cite{Gaiotto:2004ij}.
We will refer to such probe configurations as `near-horizon
microstates' in what follows. In the case of the four-dimensional
D0-D4 black hole and for large D0-charge, the relevant probe
configurations are particular bound states of D0-branes
\cite{Simons:2004nm,Gaiotto:2004pc}. An attractive feature of the
approach is that it gives insight into the physical mechanism
behind the finite number of quantum states per unit horizon area
of the black hole. The D0-brane probes experience a magnetic field
on the internal space, which effectively divides the horizon into
cells, each cell corresponding to a lowest Landau level ground
state of the D0-brane probe mechanics\footnote{See
\cite{Costa:1997zy} for an earlier application of Landau levels in
black hole physics.}. A refined version of this approach was
proposed in \cite{Denef:2007yt}, where it was argued that the
probe brane quantum mechanics arises from  the moduli space
quantization of a multicentered solution carrying the same charges
as the black hole. The mirror type IIB black hole case was
considered in \cite{Aspinwall:2006yk}, and further related work
appears in \cite{Kim:2005yb,Das:2005za}.

In this work, we will generalize the near-horizon microstate approach
to the case of the five-dimensional `D1-D5-P' black hole in type IIB carrying  wrapped D1-brane and D5-brane charges
and momentum.
We will construct near-horizon probes which are particular bound states of D1-branes and give an accounting of the
entropy  for large D1-charge. As in the four-dimensional examples, the degeneracy comes from counting lowest Landau
levels in a magnetic field on the moduli space of the probe branes.

Our motivation for transposing the approach of \cite{Gaiotto:2004ij} to the D1-D5-P black hole of \cite{Strominger:1996sh,Callan:1996dv} is twofold.
Firstly, in the D1-D5-P black hole  there is a  detailed understanding of the microscopic physics in terms of a dual conformal
field theory (see \cite{David:2002wn} for a review).
Hence we hope it will provide a good setting to address aspects of the near-horizon microstate approach which
are not fully understood at present, such as incorporating subleading corrections to the entropy.
A second motivation is that D1-D5 black holes provide the setting for another approach to
black hole physics that was  advocated by Lunin and Mathur  \cite{Lunin:2001jy}. In this approach, black hole  `hair' is represented by
 a family of nonsingular, horizonless classical supergravity solutions. For the D1-D5-P black hole, a subset of the
microstate geometries is known \cite{Mathur:2003hj}. It therefore seems a good starting point for trying to make contact
between both approaches.

We will now summarize the contents of this paper. We start by considering a four-dimensional
BPS black hole which carries 4 charges $n,w,N,W$ with metric
\be ds_4^2 = - ( H_n H_w H_N H_W )^{-1/2} dt^2 + ( H_n H_w H_N H_W )^{1/2} (dr^2 + r^2 d\O_2^2) \label{4dbh}\ee
and Bekenstein-Hawking entropy
\be S_4= 2\p \sqrt{n w N W } \label{entr4d}.\ee
We will consider two different embeddings, related by a U-duality transformation, of such a black hole in
toroidally compactified type II string theory.

In the first
duality frame, referred to as `frame A' and described in section \ref{frameA}, the charges correspond to
D0-branes and D4-branes wrapping internal cycles. This is the setting of
 \cite{Gaiotto:2004ij}. The near-horizon microstates are bound states of D0-branes that have expanded, through a form of the Myers effect
\cite{Myers:1999ps},
 to form a D2-brane wrapping the horizon $S^2$. We review this solution and its symmetry properties in
section \ref{nearhorprobeA}.
Quantum mechanically, these D0-branes are described by a superconformal mechanics with $SU(1,1|\, 2)$ symmetry.
Because they experience a magnetic field from the D4-branes in the background, as reviewed in section \ref{landauA},
their supersymmetric ground states have a large lowest Landau level degeneracy, which accounts for the entropy
(\ref{entr4d}).

The second duality frame we will consider, referred to as `frame B' and described in section \ref{dualchain}, is an
embedding as a `D1-D5-P-KK' black hole \cite{Johnson:1996ga,Constable:2000dj} in
type IIB where the charges come from  D1-branes, D5-branes, momentum and Kaluza-Klein (KK) monopole charge. The momentum and
KK monopole charges produce nontrivial fibrations for two internal circles, such that the near-horizon geometry
has a component that is locally $AdS_3$ times a squashed three-sphere $S^3/Z_W$. Following the fate of the
near-horizon microstate probes under
the U-duality  to frame B, we find that the relevant configuration is a bound state of D1-branes that has expanded
to form a Kaluza-Klein monopole that wraps the  horizon $S^3/Z_W$. We construct such a configuration explicitly as
a solution of the Kaluza-Klein monopole worldvolume action \cite{Bergshoeff:1998ef,Eyras:1998hn,Janssen:2007dm} (reviewed in section \ref{kkreview}) and show that
it has the expected symmetry properties in section \ref{KKsol}. In section \ref{landauB} we show that the solution
has a moduli space dynamics which includes a magnetic field, again reproducing the entropy from the counting of
lowest Landau levels.

In section \ref{d1d5conn}, we argue that similar Kaluza-Klein monopole solutions play the role of near-horizon
microstates for the five dimensional black hole
with D1-D5 and momentum charges and metric
\be ds_5^2 = - ( H_n H_w H_N )^{-2/3} dt^2 + ( H_n H_w H_N  )^{1/3} (dr^2 + r^2 d\O_3^2) \label{5dbh}.\ee
The argument  uses a version of the 4D-5D connection \cite{Gaiotto:2005gf,Gaiotto:2005xt,Behrndt:2005he}: by decompactifying one of the internal circles, the D1-D5-P-KK
black hole considered above lifts to a five-dimensional D1-D5-P black hole in the center of an orbifold space
$R^4/Z_W$. Since the size of the
decompactified circle is a fixed scalar, the near-horizon geometry  does not change under the decompactification,
and the relevant near-horizon probes are again bound states of D1-branes  expanded
to form a Kaluza-Klein monopole.
The special case $W=1$ gives the black hole in flat space (\ref{5dbh}), and  counting the lowest Landau level degeneracy
reproduces its entropy
\be S_5= 2\p \sqrt{n w N }\label{entr5d} .\ee

\section{The D0-D4 black hole in type IIA}
In this section we review some aspects of the near-horizon microstate approach for the four-dimensional
D0-D4 black hole in type IIA \cite{Gaiotto:2004ij}.
For simplicity, we consider a toroidal $N=8$ compactification throughout this paper, but the arguments could be
repeated for  the case of a $N=4$ compactification on $T^2 \times K_3$.

\subsection{Background}\label{frameA}
We first consider type IIA compactified on a rectangular six-torus $T^6$ which
we regard as a product of two circles $S^1,  \tilde S^2$ and two tori $T^2, \tilde T^2$.
We embed the  4-dimensional $1/8$ BPS  black hole of (\ref{4dbh}) with charges $n,w,N,W$
as a configuration consisting of D0-branes and D4-branes wrapping internal cycles as follows:
\begin{center}
\begin{tabular}{|ll|}\hline
(frame A) \qquad & $n$ D0-branes\\
& $w$ D4-branes wrapped
on $T^2 \times\tilde T^2$\\
& $N$ D4-branes wrapped
on $S^1 \times\tilde S^1 \times T^2$\\
& $W$ D4-branes wrapped
on $S^1 \times\tilde S^1 \times\tilde T^2$\\ \hline
\end{tabular}
\end{center}
We will refer to this string theory embedding as `duality frame A' in what follows.
The 10-dimensional string  metric is
\bea
ds_{10}^2  &=&- ( H_n H_w H_N H_W )^{-1/2} dt^2 + ( H_n H_w H_N H_W )^{1/2} (dr^2 + r^2 d\O_2^2)\nonu && +
{1\over 4}\left( {H_n H_w \over H_N H_W}\right)^{1/2} \left({R^2} dx^2 + {\tilde R^2}d\tilde x^2\right) + \left( {H_n H_W \over H_w H_N}\right)^{1/2}
ds^2_{T^2} + \left( {H_n H_N \over H_w H_W}\right)^{1/2}
ds^2_{\tilde T^2}\nonumber\eea
Here, $R$ and $\tilde R$ denote the radii of $S^1$ and $\tilde S^1$ on which we have chosen coordinates $x, \tilde x$
with periodicity $4\p$.
We will work in the units  $2 \p \sqrt{\a '} =1 $, where the fundamental string and D-brane charges take the
value $2\p$. The harmonic functions are given in terms of the
quantized charges as
\be\begin{array}{lcl} H_n = 1+ {g_\infty \over 4 \p V_{T^6}} {n \over r} &\qquad&
H_w = 1+ {g_\infty \over 4\p  (2 \p R)(2 \p \tilde R)} {w \over r}\\
H_N = 1+ {g_\infty \over 4 \p V_{\tilde T^2} } {N \over r} &\qquad& H_W = 1+ {g_\infty \over 4 \p V_{T^2}} {W \over r}\end{array}\ee
The dilaton and RR gauge fields are given by
\bea
e^{\F} &=&{ g_\infty} H_n^{3/4} (H_w H_N H_W)^{-1/4} \nonu
C^{(1)} &=& -{1 \over g_\infty} \left({1\over H_n} -1 \right) dt;\qquad C^{(3)} = -{1\over 4 \p} \cos \theta d\f\wedge
\left[ w\ \o_{S^1\times\tilde S^1} + W  \o_{T^2} + N  \o_{\tilde T^2}\right]\nonumber
\eea
Here, the $\o_{\cm_2}$ are normalized  volume forms satisfying $\int_{\cm_2} \o_{\cm_2}=1$  and $g_\infty$ is the value of the string coupling at  infinity.

We will be interested in the near-horizon scaling limit of this geometry, which is obtained by
temporarily restoring $\a '$ factors and
taking
\be \a ' \to 0; \qquad \qquad   {r\over \a'},\ {R\over \sqrt{\a '}},\ {\tilde R\over  \sqrt{\a '}}
,\ {V_{T^2}\over \a'},\ {V_{\tilde T^2}\over \a'} \ {\rm fixed} \label{NHscalingA}\ee
In this limit,
the above background reduces to an $AdS_2 \times S^2 \times T^6$ attractor geometry
where the $AdS_2 \times S^2$ radius and the volumes of the tori $S^1\times\tilde S^1,\ T^2 $ and $\tilde T^2$ are fixed
in terms of the charges. Performing a coordinate
change to global $AdS_2$ coordinates
\be
r = l_A (\cosh \c \cos \t + \sinh\c ); \qquad
t = {l_A^2 \over r} \cosh\c \sin\t  \label{coordA}
\ee
as well as  a gauge transformation on $C^{(1)}$, we obtain the near-horizon geometry
\bea
ds^2_{10} &=& l_A^2\left[ - \cosh^2 \c d\t^2 + d\c^2 + d\q^2 + \sin^2\q d\f^2 \right]\nonu
&& + \sqrt{ n\over w N W}
\left[ {w\over 16 \p^2} \left({R \over \tilde R} dx^2 + {\tilde R \over R} d \tilde x^2\right) + {W\over V_{T^2}}
{ds^2_{T^2}}
  + { N\over V_{\tilde T^2}} {ds^2_{\tilde T^2}}\right]\nonu
C^{(1)} &=&  -{1\over 4 \p} \sqrt{w N W \over n} \sinh \c d\t;
\qquad C^{(3)} =  - {1\over 4\p} \cos \theta d\f \left[ w\, \o_{S^1 \times\tilde S^1}+
W\o_{T^2}+ N\o_{\tilde T^2}\right]\nonu\label{nearhorA}\eea
The $AdS_2 \times S^2$ radius $l_{A}$ is given by
\be l_{A} = {g\over 4 \p} \sqrt{w N W \over n} .\label{lA}\ee
Here, $g$ denotes the value of the string coupling in the near-horizon region.
The supergravity description is reliable as
long as $g\ll 1$ and $l_A \gg 1$.

 The near-horizon geometry preserves 8 Killing spinors,
which combine with  the $SL(2,R)$ isometry group of $AdS_2$ and the $SO(3)$ symmetry of $S^2$ into
an $SU(1,1|2)$ super-isometry group. The Killing vectors generating the $SL(2,R)$ component are given by
\bea
l_0 &=& \pa_\t\nonu
l_\pm &=& e^{\pm i\t} \left[ \tanh \c \pa_\t \mp i \pa_\c \right]\label{sl2rA}
\eea
\subsection{Horizon-wrapping membranes and their symmetries}\label{nearhorprobeA}
The near-horizon microstates that capture the entropy  of the D0-D4 black hole at large
D0-charge are particular bound states of D0-branes that have expanded to form a D2-brane, wrapping the
horizon $S^2$, through a form of
the Myers effect. These can be described as noncommutative solutions in the multi-D0-brane action
or, alternatively, as solutions of the D2-brane action with D0-brane charge dissolved on the worldvolume.
We will here focus on the latter description and describe the probe solution and its properties in more detail.

We consider a D2-brane probe in the background (\ref{nearhorA}),
wrapping the horizon $S^2$, and choose a static gauge such that the worlvolume coordinates coincide with
$\t, \theta, \f$.
Turning on woldvolume flux $\cf $ on
$S^2$ induces $Q$ units of D0-brane charge:
$$ \cf = {Q\over 4\p} \sin \theta d\theta d\f .$$
Dimensionally reducing over the two-sphere,
the Lagrangian describing the motion of such a brane reads
\be \cl = - M l_A \left[  \sqrt{ 1 + \r^2} \sqrt{\cosh^2 \c  -
\dot \c^2 } + \sinh \c \right].\label{D2lag}\ee
We have restricted attention to a D2-brane that is static  on the
$T^6$ for the time being. The parameters $M$ and $\r$ correspond to the
mass of the wrapped D2-brane and the induced D0-brane charge density on $S^2$
respectively:
\bea
M &=& 4 \p l_A^2 T_{D_2} = {g \over 2} {w NW \over n} \label{MA}\\
\r &=& {Q \over 4 \p l_A^2} = {4\p Q \over g^2} {n\over w NW} .\nonumber
\eea
The isometries (\ref{sl2rA}) of the background act as symmetries on the D2-brane worldvolume
and lead to Noether charges $L_0, L_\pm$, where $L_0$ is the canonical Hamiltonian
obtained from (\ref{D2lag}). They are given by
\bea
L_0 &=& \cosh\c \sqrt{ P_\c^2 + (M l_A)^2(1 + \r^2)} + M l_A \r \sinh\c \label{hamA}\\
L_\pm &=& e^{\pm i \t} \left[\tanh \c L_0 \pm i P_\c + {M l_A \r \over \cosh \x} \right]\label{noetherA}
\eea
These expressions are derived by varying the D2-brane action before gauge-fixing the worldvolume time coordinate,
and the last term in (\ref{noetherA}) arises because the Wess-Zumino term is only invariant up a total derivative.
From (\ref{D2lag}) or (\ref{hamA}) we see that there is a static solution where the brane
is located at
$$ \sinh \c = - \r .$$
The Noether charges (\ref{noetherA}) evaluated on this solution are \cite{Das:2005za}
\bea
L_0 &=& M l_A \label{BPSA}\\
L_\pm &=& 0
\eea
Hence the solution is `primary' and, in addition, invariant under conformal boosts:
\be K= L_+ + L_-=0 .\label{confboost}\ee
The supersymmetry properties of this solution were analyzed in \cite{Simons:2004nm}, where it was shown to
preserve half of the near-horizon supersymmetries. Since it is static with respect to global time $\t$ instead of
Poincar\'{e} time $t$, it breaks all of the Poincar\'{e} supersymmetries  that extend to the asymptotically flat region.
Such branes are necessarily bound to the near-horizon region
and have an energy barrier preventing them to escape to asymptotic infinity.

\subsection{Landau levels on moduli space and microstate counting}\label{landauA}
A important property of the horizon-wrapping membranes is that they experience a magnetic field, induced by the D4-brane
charges in the background, on their moduli
space. Due to this fact, the supersymmetric ground states in the quantized theory  have a large lowest Landau level
degeneracy which accounts for
the black hole entropy, as we shall presently review.

The energy of the probe solutions  is independent of the position of the probe on $T^6$. Hence these positions
are  bosonic moduli of our solution and the moduli space $\cm$ equals $T^6$.
The moduli space mechanics is that of a particle moving on $T^6$, with  kinetic
terms come from expanding the
Born-Infeld action. The  particle also couples (with charge $2\p$) to a magnetic field which comes from the  Wess-Zumino
coupling $\int C^{(3)}$ to the D4-branes in the background. From the expression (\ref{nearhorA}), we see that
this term gives
rise to a magnetic field with field strength
\be
F_\cm =  w\,\o_{S^1 \times\tilde S^1}+
 W\o_{T^2}+   N\o_{\tilde T^2}\label{magnfieldA}
\ee
Hence the particle moves in a magnetic field with $w$ units of  flux through $S^1\times\tilde S^1$,
$W$ units of flux through $T^2$ and $N$ units of flux through $\tilde T^2$.

The full quantum mechanical theory describing the low-energy dynamics of the horizon-wrapping membrane was constructed
in \cite{Gaiotto:2004pc}. The theory realizes the superconformal algebra $SU(1,1|\, 2)$ with a central charge $M l_A$, with $M$ given in
(\ref{MA}). The corresponding
BPS bound is saturated by chiral primary states which satisfy (\ref{BPSA}). In \cite{Gaiotto:2004ij}, it was shown that these
chiral primaries are in one-to-one correspondence with lowest Landau levels in the magnetic field $F_\cm$.
More precisely, by using a standard representation for the fermionic zero modes on differential forms
in moduli space, it was shown that chiral primaries are represented by harmonic forms $h$ with respect to a covariant
derivative $\bar D$:
\be \bar D h = \bar D^\dagger h = 0\label{coh}\ee
where $\bar D = \bar \pa + \bar A_{\cm}$ and $\bar A_{\cm}$ is a holomorphic connection for $F_{\cm}$.
Even forms represent bosons while odd forms represent fermions.
The number of solutions to (\ref{coh}) could be computed by index theory as in \cite{Gaiotto:2004ij} but, in the simple
toroidal case we are considering, one can also enumerate the solutions explicitly.
On a single two-torus $\cn$ units of magnetic flux, there are harmonic (0,0) forms $\Psi_J$ and harmonic  (1,0) forms
$\Psi_J dz$.
The  $\Psi_J$, where $J$ is an angular momentum quantum number running from 0 to $\cn -1$, are lowest
Landau level wavefunctions on the torus whose explicit form can be found e.g. in \cite{Haldane}.
In the case at hand, we have a product space of three tori with magnetic field, and we get a total
of $4 w N W$ bosonic and $4 w N W$ fermionic solutions.

It is important to observe that the number of chiral primaries does not depend on the background D0-charge
$n$, and for the purpose of state counting we can imagine all the D0-charge to be carried by the probe branes.
The total D0-charge $n$ can then be divided in many ways into clusters of D0-branes forming D2 bound states
considered above. Counting such multiparticle  chiral primaries is equivalent to counting the degeneracy in a
CFT with $4 w N W$ bosons and $4 w N W$ fermions at level $n$. The central charge $c$ is  $6 w N W$  and the
degeneracy $D(n)$ at large $n$ is given by the Cardy formula
$$ \ln D(n) = 2 \p \sqrt{n c /6} = 2\p \sqrt{ nwNW} $$
in agreement with the macroscopic entropy (\ref{entr4d}).\\
We end this section with some remarks:
\begin{itemize}
\item The above picture, where we counted bound states of D0-branes, was valid for
black holes which are `mostly made up out of D0-branes' where the D0-charge $n$ is parametrically
larger than the D4-brane charges. As a result, the calculation does not capture corrections to the entropy
subleading in $n$, which can easily be seen from the fact that the full D0-brane partition function is not U-duality invariant.
\item We should also remark that the above calculation is not a fully controlled approximation:
in the limit of parametrically large $n$ where we did the microstate counting, we see from $(\ref{lA})$ that it
is not possible to keep both the string coupling small while keeping the $S^2$ radius large in string units.
 Strong coupling problems of this type are typically resolved by going to a different duality frame where the approximations
are under control; this will be the case for the U-dual description we will consider in the next section.
\item We also observe that there is a similar picture of near-horizon microstates for black holes
which are `mostly made up out of D4 branes', i.e. where one of the D4-charges is parametrically larger than the other
charges. This situation is T-dual to the case considered above. The near-horizon probe solutions are now D6-branes which wrap the horizon $S^2$ and a four cycle, with
worldvolume flux on the $S^2$. The moduli space is again a $T^6$, spanned by  two transverse directions
and four Wilson lines, and magnetic fields on moduli space are produced by the couplings
$\int C^{(3)}\wedge \cf\wedge \cf$ and $\int C^{(7)}$.

\end{itemize}

\section{U-duality and the D1-D5-P-KK black hole}
\subsection{Duality chain}\label{dualchain}
In this section we will describe an embedding of the 4-dimensional black hole (\ref{4dbh}) as
a `D1-D5-P-KK' black hole solution in type IIB, with the charges $n,w,N,W$ corresponding to D1, D5, momentum and
Kaluza-Klein monopole
charge respectively. This solution can be obtained from the one in duality frame A
in the previous section through a U-duality transformation of the form $TST$:
\TABLE{}{
\begin{center}
\begin{tabular}{lclclcl}
IIA & & IIB &  & IIB & &IIB \\
D0 & & D3 & &   D3 & &   D1($ S^1$) \\
 D4($T^2,\tilde T^2$) & T ($\tilde S^1, T^2$) & D3& S &  D3&T ($S^1 , \tilde S^1 , T^2$)&  D5($S^1, T^2,\tilde T^2$)\\
 D4($S^1 ,\tilde S^1 , T^2$) & $\longrightarrow$ &  D1& $\longrightarrow$ & F1&$\longrightarrow$&  P($S^1$)\\
 D4($S^1 ,\tilde S^1 , T^2$) & &  D5 & &  NS5& & KK($S^1 ,\tilde S^1_{TN},T^2, \tilde T^2$)\label{Udual}
\end{tabular}
\end{center}}The final configuration,
which we will refer to as `duality frame B', is
\begin{center}
\begin{tabular}{|ll|}\hline
(frame B) & $n$ D1-branes wrapped on $ S^1$\\
& $w$ D5-branes wrapped
on $S^1\times T^2\times\tilde T^2$\\
& $N$ units of momentum on
on $S^1$\\
& $W$ KK-monopoles, Taub-NUT direction $\tilde S^1$, wrapped
on $S^1  \times T^2\times \tilde T^2$\\\hline
\end{tabular}
\end{center}

In this duality frame, the $S^1$ and $\tilde S^1$ circles are fibred nontrivially due to the momentum and KK monopole
charges. The 10-dimensional metric is
\be\begin{array}{l}
ds_{10}^2  = ( H_n H_w )^{-1/2} \left[ - {1\over H_N} dt^2 + H_N \left( {R \over 2} d x - ({1/H_N}-1)dt  \right)^2\right]
 \\
+ ( H_n H_w  )^{1/2} \left[H_W (dr^2 + r^2 d\O_2^2) + {\tilde R^2 \over4 H_W} (  d \tilde x - W\cos \theta d\f )^2 \right]
 + ( {H_n / H_w })^{1/2} (ds^2_{T^2} + ds^2_{\tilde T^2})
\label{metricB}\end{array}\ee
The harmonic functions are now given by
\be\begin{array}{lcl} H_n = 1+ {g_\infty \over 4 \p (2 \p \tilde R) V_{T^2}V_{\tilde T^2}} {n \over r} &\qquad&
 H_w = 1+ {g_\infty \over 4\p  (2 \p  \tilde R)} {w \over r}\\
H_N = 1+ {g_\infty^2 \over 4 \p (2\p R) V_{T^6} } {N \over r}  &\qquad& H_W = 1+ {\tilde R \over 2} {W \over r}.\end{array}\ee
For the dilaton and RR fields, we have
\bea
e^{\F} &=&{ g_\infty} H_n^{1/2} H_w^{-1/2} \nonu
C^{(2)} &=& -{R \over 2 g_\infty} \left({1/ H_n} -1 \right) dt\wedge dx- {w \over 16\p^2} \cos \theta d\f\wedge d\tilde x\label{15pK}
\eea

We now  take a near-horizon scaling limit that matches the one we considered in frame A (\ref{NHscalingA}), as well as
a rescaling of $t$ and a coordinate change (\ref{coordA}). One of the gauge transformations we performed in
frame A now becomes a shift of the coordinate $x$, leading to a new variable by $x'$ with period $4\p$.
We obtain an attractor geometry where, this time, the fixed moduli are the radii of
${S^1}$ and $\tilde S^1$ and the volume of the 4-torus $T^2 \times\tilde T^2$:
\bea
ds^2_{10} &=& l_B^2\left[ - \cosh^2 \c d\t^2 + d\c^2  +  d\q^2 + \sin^2\q d\f^2 \right]+ k^2( dx' + A)^2
+ \tilde k^2 (d\tilde x + \tilde A)^2\nonu
&&+ \sqrt{n \over w} {ds^2_{T^2} + ds^2_{\tilde T^2}\over \sqrt{V_{T^2} V_{\tilde T^2}}}\nonu
C^{(2)} &=& {1\over g} ( k^2 A \wedge d x' + \tilde k^2 \tilde A \wedge d \tilde x )\label{nearhorB}
\eea
with the Kaluza-Klein scalars  $k,\tilde k$ and  one-forms  $A,\tilde A$ given by
\be k^2 = l_B^2 {N\over nwW}; \qquad \tilde k^2 =  {l_B^2\over W^2};\qquad  A = - \sqrt{n w W\over N} \sinh \c d \t;
\qquad \tilde A = - W \cos \theta d \f. \ee
The  radius $l_{B}$ is given by
\be l_{B} = {\sqrt{g} \over 4 \p} \sqrt{w W} \label{lB}\ee
where $g$ is the string coupling in the  near-horizon region.
We can trust the supergravity description  as
long as $g\ll 1$ and $g w W\gg 1$.

In (\ref{nearhorB}), the circles ${S^1}$ and $\tilde S^1$ are `Hopf'-fibered over $AdS_2$ and $S^2$ respectively so as
to form a space
which is locally $AdS_3 \times S^3$ with curvature radii $l_{AdS_3} = l_{S^3}= 2 l_B$. Due to the compactness of
 $x'$ and the KK monopole charge $W$ (when $W>1$)
however, the space is not globally $AdS_3 \times S^3$, but rather the product of `squashed' $AdS_3$ with the squashed
three-sphere $S^3/Z_W$ \cite{Duff:1998cr}.  The squashing preserves the
left-moving isometry group $SL(2,R)_L  \times SO(3)_L$, combining with fermionic generators into an $SU(1,1|\, 2)$
supergroup,
while the right-moving $SL(2,R)_R  \times SO(3)_R$ is broken down
to two $U(1)$'s which act as translations of $x'$ and $\tilde x$.
Hence we find the same  super-isometry group as in frame A, as of course we should.
The Killing vectors generating $SL(2,R)_L$ are given by
\bea
l_0 &=& \pa_\t\nonu
l_\pm &=& e^{\pm i\t} \left[ \tanh \c \pa_\t \mp i \pa_\c - {1\over \cosh\c }\pa_{x'}\right]\label{sl2rB}
\eea

Having discussed the duality transformation relating the D0-D4 black hole to the D1-D5-P-KK black hole,
we will now apply the same dualities to the near-horizon microstate probes of frame A in order
to find  microstate probes in frame B.
Under the U-duality transforming frame A into frame B, a D2-brane wrapping $S^2$ in frame A transforms as:
\TABLE{}{
\begin{center}
\begin{tabular}{lclclcl}
IIA & T ($\tilde S^1, T^2$) & IIB & S & IIB &T ($S^1 , \tilde S^1 , T^2$)  &IIB \\
D2($S^2$) & $\longrightarrow$ &  D5
& $\longrightarrow$ &   NS5& $\longrightarrow$ & KK ($S^1_{TN},S^3/Z_W , T^2 $)\\ \label{Udualprobe}
\end{tabular}
\end{center}}
The first arrow is essentially  mirror symmetry, and leads to the near-horizon probe picture discussed in \cite{Aspinwall:2006yk}.
The final probe configuration in frame B is a Kaluza-Klein (KK) monopole wrapped on the near-horizon $S^3/Z_W$ as well as on $T^2$
and whose Taub-NUT direction is along $S^1$. The configuration in frame A also carried D0-brane charge $Q$
induced by  worldvolume flux on $S^2$. The KK-monopole probe in frame B similarly has an appropriate woldvolume field
turned on so as to induce D1-brane charge along $S^1$. We will construct such a solution explicitly in section \ref{KKsol}
from the worlvolume action of the KK monopole,
and check that it carries the same Noether charges as its counterpart in frame A. In section \ref{landauB}, we will
show that its moduli space mechanics includes magnetic fields of the
correct magnitude.

\subsection{The worldvolume action for a KK monopole}\label{kkreview}
Let us first review some facts about the effective worldvolume description of KK-monopoles. The
worldvolume theory of a KK-monopole in type IIB  is a $(2,0)$ theory in 5+1 dimensions, and the
 collective coordinates organize themselves into a tensor multiplet \cite{Sen:1997js}.
The worldvolume dynamics, which is determined by dualities relating the KK-monopole to other branes \cite{Eyras:1998hn}, cannot be
captured
by a covariant action due to the selfduality condition on the tensor field. To avoid this difficulty,
we will make use of the observation of
\cite{Janssen:2007dm} that, after dimensional reduction to 4+1 dimensions, the tensor multiplet reduces to a $(1,1)$ vector multiplet
which can be described by a covariant action. Therefore, if we consider KK monopole which is wrapped on at least one
compact direction, we can use a dimensionally reduced 4+1 dimensional action, which can be obtained by T-dualizing
the action for the type IIA KK-monopole constructed in \cite{Bergshoeff:1998ef} along a worldvolume direction.

The wrapped KK-monopole action thus obtained can be formulated in spacetimes which have two compact isometry directions:
the first one, which we will call $k^\m$, denotes the Taub-NUT circle of the monopole, while the second one, $\tilde k^\m$,
is the circle on which the monopole is wrapped.
The field content is summarized in  the table below  and consists of three scalars $X^i$, describing
transverse fluctuations,
two zero-forms  $\o^{(0)}, \tilde \o^{(0)}$ (with field strengths $\cg^{(1)}, \tilde \cg^{(1)}$)
which source fundamental and D-string charge along the Taub-NUT direction $k^\m$, and a one-form $\ca^{(1)}$ (with
field strength $\cf^{(2)}$).
\TABLE{}{\begin{center}
\begin{tabular}{|c|c|}\hline
worldvolume field & field strength   \\ \hline \hline
$X^i$ & -\\ \hline
$\o^{(0)}$ & $\cg^{(1)}$\\ \hline
$\tilde \o^{(0)}$ & $ \tilde \cg^{(1)}$ \\ \hline
$\ca^{(1)}$ & $\cf^{(2)}$ \\
 \hline
\end{tabular}
\end{center}}

We now place a KK-monopole probe in the background (\ref{nearhorB}), taking the Taub-NUT direction
to be along the circle $S^1$: $k^\m = (\pa_{x'})^\m$. It will be convenient to choose the wrapping direction
to be along the $\tilde S^1$ circle so that $\tilde k^\m = (\pa_{\tilde x})^\m$. The wrapped KK-monopole action in
this background reduces to
\bea
S &=& -\t_{KK} \int d^5 \s\, k^2 \tilde k e^{-2\F} \sqrt{-\det ( P[ \tilde G]_{ab} + k^{-2}\cg^{(1)}_a\cg^{(1)}_b +
 e^{2\F}k^{-2}\tilde \cg^{(1)}_a\tilde \cg^{(1)}_b - e^\F  k^{-1} {\tilde k}^{-1} \cf^{(2)}_{ab})}\nonu
&&+ \t_{KK} \int \left[ P[i_{\tilde k} i_{k}  N^{(7)}] + \half P[\tilde A] \wedge \cf^{(2)} \wedge \cf^{(2)}
+ P[A] \wedge \cf^{(2)} \wedge  \cg^{(1)}\wedge \tilde \cg^{(1)}\right]\label{KKaction}
\eea
Here, we have denoted pullbacks by $P[\ldots ]$, while $\tilde G_{\m\n}$ is the metric on the 8-dimensional
 base space over which
the $S^1$ and $\tilde S^1$ circles are fibered:
$$ \tilde G_{\m\n} = G_{\m\n}  - {k_\m k_\n \over k^2} - {\tilde k_\m \tilde k_\n \over \tilde k^2}.$$
The 7-form $N^{(7)}$ is a gauge potential for the 8-form field strength
which is dual to the KK one-form $A$, with
$$ i_{\tilde k} i_{k}  N^{(7)} = { 2 \p P \over \t_{KK}}\  \o_{T^2}
\wedge \o_{S^2}\wedge  {\tilde y^1 d\tilde y^2\over V_{\tilde T^2}}. $$
The worldvolume field strengths entering in (\ref{KKaction}) are
\bea
\cg^{(1)} &=& d \o^{(0)} \nonu
\tilde \cg^{(1)} &=& d \tilde \o^{(0)} + P[i_{k} C^{(2)}] \label{D1coupling}\\
\cf^{(2)} &=&  dA^{(1)}/(4 \p)^2 - P[i_{\tilde k} C^{(2)}]\wedge \cg^{(1)} .\label{wvfs}
\eea
The normalization constant $\t_{KK}$ takes the value $\t_{KK} = 8 (2\p)^4$ in our units, and is related to the physical
tension $T_{KK}$ as $T_{KK} = \t_{KK} k^2 \tilde k / g^2$.

\subsection{Horizon-wrapping KK-monopoles and their symmetries}\label{KKsol}
We shall now explicitly construct the KK-monopole probe solutions which will play the role of near-horizon microstates
of the D1-D5-P-KK black hole, and show that they have the same symmetry properies as their counterparts in frame A.
Choosing coordinates $y^1, y^2$ on  $T^2$ and $\tilde y^1, \tilde y^2$ on  $\tilde T^2$,
we will work in a static gauge where the worldvolume coordinates are identified with $\t,\theta,\f,y^1,y^2$.
As discussed in section \ref{dualchain}, we want to consider a KK-monopole that carries induced D1-brane charge by
turning on appropriate worldvolume fields. Such a solution can  be interpreted as a bound state of D1-branes that
have expanded into a KK monopole through the Myers effect.
From the relation (\ref{D1coupling}) we see that turning on time-dependent $\tilde \o^{(0)}$ sources D1-brane charge along $S^1$,
so that the conserved momentum conjugate to $\tilde \o^{(0)}$ will be proportional to the induced D1-brane charge.

The Lagrangian for such a KK monopole, dimensionally reduced over $S^2 \times T^2$, is given by
\be
\cl = - M l_B \sqrt{ \cosh^2 \c - (\b \dot{\tilde \o}^{(0)} - \sinh \c)^2}\label{kkmonlagB}
\ee
We have restricted attention to a static probe on $\tilde T^2$ with constant gauge fields $\o^{(0)}, \ca^{(1)}$.
The constants $M$ (representing the mass of the wrapped KK-monopole) and $\b$ are given by
\be
M = T_{KK} 4 \p l_B^2 \sqrt{ n\over w} \sqrt{ V_{T^2} \over V_{\tilde T^2}};\qquad \qquad
\b = {g \over k^2}\left( {N \over nw W}\right)^{1/2}
\ee
The momentum conjugate to $\tilde \o^{(0)}$ is related to the induced D1-brane charge $Q$ as
\be P_{\tilde \o^{(0)}} = 2 \p Q.\label{indD1charge}\ee

We observe from the form of (\ref{kkmonlagB}) that the background fields in (\ref{nearhorB}) have
conspired to produce a Lagrangian  describing the  motion of a particle
on a locally  $AdS_3$ space, with $\tilde \o^{(0)}$ playing the role of the Hopf fibre coordinate.
Hence  we can easily find the $SL(2,R)_L$ Noether charges $L_0, L_\pm$ from (\ref{sl2rB}):
\bea
L_0 &=& \cosh\c \sqrt{ P_\c^2 + (M l_B)^2(1 + \r^2)} + M l_B \r \sinh\c \label{hamB}\\
L_\pm &=& e^{\pm i \t} \left[\tanh \c L_0 \pm i P_\c + {M l_B \r \over \cosh \x} \right].
\eea
Here, $L_0$ is the canonical Hamiltonian obtained from (\ref{kkmonlagB}).
To derive these expressions, we have used (\ref{indD1charge}) and defined $\r$ as $\r = { 2 \p Q / M l_B \b}$.
With these definitions, the Noether charges take precisely the same form as in frame A (\ref{noetherA}). Again, there
is a  static solution which now represents
a wrapped KK-monopole located at \be \sinh \c = - \r \label{kksolB}.\ee
It has Noether charges $L_0 = M l_B  , L_\pm = 0$ and is the sought-after configuration U-dual
to the horizon-wrapping membrane in frame A.

As the $\kappa$-symmetry transformations of the KK-monopole action are not known at present, it is not possible
to directly verify the preserved supersymmetries of the solution. U-duality predicts
that it should have the same supersymmetry properties as its counterpart in frame A, namely
preserving half of the supersymmetries while
breaking all Poincar\'{e} supersymmetries.

\subsection{Moduli space dynamics and state counting}\label{landauB}
We will now consider the moduli space dynamics of the probe solution considered above. As in frame A, we will see that
our probes experience
a magnetic field on moduli space and that the lowest Landau level degeneracy accounts for the Bekenstein-Hawking entropy.
Due to the complexity of the action (\ref{KKaction}), the analysis will be more involved than
in frame A. We will see that the magnetic field on moduli space now arises both from Born-Infeld and
Wess-Zumino terms in the action (\ref{KKaction}).

The energy of the above solutions is independent of the constant values of the worldvolume fields $\o^{(0)}, \ca^{(1)}, \tilde y^1,
\tilde y^2$, hence these will give rise to the  moduli of the solution. The moduli space mechanics is obtained in a standard manner
by expanding the action around the solution (\ref{kksolB}) to quadratic order in the fields $\o^{(0)}, \ca^{(1)}, \tilde y^1,
\tilde y^2$ and dimensionally reducing to 0+1 dimensions.
The quadratic action is
\bea
S_2 &=& - \t_{KK} \int \Big[ k^2 \tilde k e^{-2 \F}\sqrt{ n\over w V_{T^2}V_{\tilde T^2}} (d \tilde y^1 \wedge \star
d \tilde y^1  + d \tilde y^2 \wedge \star
d \tilde y^2 ) + \half \tilde k e^{-2\F} d \o^{(0)} \wedge \star d \o^{(0)}\nonu
&& +{1\over 4 \tilde k} \cf^{(2)} \wedge \star \cf^{(2)}\Big]
 + \t_{KK} \int \Big[ P[i_{\tilde k} i_k N^{(7)}] + \half
P[\tilde A] \cf^{(2)} \wedge \cf^{(2)}\Big].\label{kwadraction}\eea
where \be \cf^{(2)} =   {dA^{(1)}\over (4 \p)^2} + {\tilde k^2\over g}P[\tilde A]\wedge d\o^{(0)}.\nonumber\ee
The Hodge $\star$ is to be taken with respect to the worldvolume metric
$$ ds^2_{wv} = l_B^2 ( - d\t^2 + d\theta^2 + \sin^2\theta d\f^2) + \sqrt{n \over w V_{T^2} V_{\tilde T^2}}
\left( (dy^1)^2+(dy^2)^2 \right) .$$
Note that the kinetic terms in (\ref{kwadraction}) are not diagonal due to the mixing between $\ca^{(1)}$ and $\o^{(0)}$.
We will now perform the dimensional reduction along with field redefinitions so
 as  to obtain diagonal kinetic terms in 0+1 dimensions.

First, we dimensionally reduce over the $T^2$ directions $y^1,y^2$ to three
dimensions. The reduction of the field $\ca^{(1)}$ gives
two Wilson lines $w^1, w^2$ from the components along $y^1, y^2$ and a gauge field $\ca'^{(1)}$ with curvature $\cf'^{(2)} = d \ca'^{(1)}$.
Next, we dualize $\ca'^{(1)}$ into a zero-form $\tilde \ca^{(0)}$ by adding a term $\int d \tilde \ca^{(0)}\wedge \cf'^{(2)}$
to the action and integrating over $\cf'^{(2)}$. This produces a kinetic term for  $\tilde \ca^{(0)}$ and, because
of the mixing terms in (\ref{kwadraction}), an
additional  Wess-Zumino type term $w / W \int  P[\tilde A]\wedge d\o^{(0)} \wedge d \tilde \ca^{(0)}$.
After partial integrations and a further reduction over $S^2$, we obtain a particle action on a rectangular six-torus with
coordinates $\o^{(0)}, \tilde \ca^{(0)},  w^1, w^2, \tilde y^1, \tilde y^2$ in a magnetic field. The magnetic field strength is
\be F_{\cm} =  w\, 2 d {\o^{(0)} \wedge d \tilde \ca^{(0)} }+ W V_{T^2}d w^1 \wedge d w^2 +
{N \over V_{\tilde T^2} } d \tilde y^1 \wedge d\tilde y^2. \label{magnfieldB}\ee
Taking into account the periodicities of the moduli space coordinates\footnote{The periodicity
of $\o^{(0)}$ is $1/4 \p$, while $\ca^{(0)}$ has period $2\p$, and $w^1,w^2$ have the inverse periodicities of $y^1,y^1$.},
we again find
three tori with $w$, $W$ and $N$ units of magnetic flux respectively.
As explained in section \ref{landauA}, the counting of chiral primary  states of the
KK-monopole theory reduces to counting lowest Landau level degeneracies and reproduces the entropy (\ref{entr4d}).

\section{4D-5D connection and the D1-D5-P black hole}\label{d1d5conn}
We now discuss the relevance of the probe solutions constructed above to the description of
near-horizon microstates in five-dimensional black holes.
The reason for this is that the D1-D5-P-KK background (\ref{15pK}) lends itself to a version of the
4D-5D connection which was also at the basis of the earlier work
\cite{Johnson:1996ga,Constable:2000dj,Cvetic:1998hg} and which we will outline here.

So far, we have assumed the radius $\tilde R$ of the $\tilde S^1$ circle to be small compared to the
size $l_B$ of the black hole. In this regime, the appropriate picture is that of
a black hole in four dimensions (\ref{4dbh}). We now vary the radius to the regime where $\tilde R\gg l_B$, where
the geometry effectively looks five-dimensional and describes a five-dimensional black hole with D1-D5-P
charges ($n,w,N$), placed at the center of a Taub-NUT space with NUT charge $W$.
The relevant limit to describe this
 five-dimensional regime is to take the decompactification limit
keeping $\tilde R r$ fixed:
\be \tilde R \to \infty; \qquad \qquad \tilde r^2 \equiv 2 \tilde R r \ {\rm fixed}\label{decomp}.\ee
The background (\ref{15pK}) becomes
\be\begin{array}{l}
ds_{10}^2  = ( H_n H_w )^{-1/2} \left[ - {1\over H_N} dt^2 + H_N \left( {R \over 2} d x - ({1/H_N}-1)dt  \right)^2\right]
 \\
+ W ( H_n H_w  )^{1/2} \left[ d\tilde r^2 + {1\over 4}\tilde r^2\left( d\O_2^2 + ({1\over W}  d \tilde x - \cos \theta d\f )^2
\right)\right]
 + ( {H_n / H_w })^{1/2} (ds^2_{T^2} + ds^2_{\tilde T^2})
\label{metric5d}\end{array}\ee
and the harmonic functions are the correct ones for objects in five noncompact dimensions:
\be\begin{array}{lcl} H_n = 1+ {g_\infty \over 4\p^2  V_{T^2}V_{\tilde T^2}} {n \over \tilde r^2} &\qquad&
 H_w = 1+ {g_\infty \over 4\p^2  } {w \over \tilde r^2}\\
H_N = 1+ {g_\infty^2 \over 4 \p^2 (2\p R)^2 V_{T^2}V_{\tilde T^2}  } {N \over \tilde r^2} . &\qquad& \end{array}\ee
For the dilaton and RR fields, we have
\bea
e^{\F} &=&{ g_\infty} H_n^{1/2} H_w^{-1/2} \nonu
C^{(2)} &=& -{R \over  2 g_\infty} \left({1/ H_n} -1 \right) dt\wedge dx- {w \over 16 \p^2}
\cos \theta d\f\wedge d\tilde x\label{dilRR5d}
\eea
In five dimensions, the metric in the Einstein frame reads
\be ds_5^2 = - ( H_n H_w H_N )^{-2/3} dt^2 + ( H_n H_w H_N  )^{1/3} W \left[d\tilde r^2 + {1\over 4}\tilde r^2
\left( d\O_2^2 + ({1\over W}  d \tilde x - \cos \theta d\f )^2 \right)\right].\label{bh5dorb}\ee
This metric describes a three-charge black hole placed in an orbifold space $R^4/Z_W$. The Bekenstein-Hawking
entropy is
\be S_5= 2\p \sqrt{n w N W}\label{entr5dorb} .\ee
The special case $W=1$ yields the five-dimensional black hole in flat space (\ref{5dbh}).

The 4D-5D connection described above leads to an explicit construction of the near-horizon
microstates of the 5D black holes (\ref{bh5dorb}). Since the modulus $\tilde R$ is, as we have seen, a fixed scalar, the
near-horizon
geometry reduces to (\ref{nearhorB}) for any value of $\tilde R$. In particular,
starting from (\ref{metric5d}),(\ref{dilRR5d}) and taking the limit
\be \a ' \to 0; \qquad \qquad
{\tilde r^2 \over {\a'}^{3/2}},\ {R \over \sqrt{\a '}},\  {V_{T^2}\over \a'}, {V_{\tilde T^2}\over \a'} \ {\rm fixed}\label{NHscaling5D}\ee
 and making
similar coordinate changes as before, we obtain precisely the same near-horizon geometry  (\ref{nearhorB}) as in frame B.
Hence the construction of the near-horizon
microstates can be taken over from section \ref{landauB}. They are again given by horizon-wrapping KK-monopoles,
whose moduli space dynamics contains a magnetic field (\ref{magnfieldB}).
As mentioned above, we are particularly interested
in the case $W=1$ describing the D1-D5-P black hole in flat space.
The counting of lowest Landau degeneracies  involves solving the harmonic equation (\ref{coh}) on a product of two tori with magnetic fluxes
$w$ and $N$ and a two-torus with one unit of magnetic flux
coming from the Hopf bundle on $S^2$. The construction of the harmonic forms in section \ref{landauA} can be applied
in this case and, proceeding as described there, the microscopic counting reproduces the Bekenstein-Hawking entropy (\ref{entr5d}).

We end with some further remarks:
\begin{itemize}
\item We should remark that the  near-horizon scaling limit (\ref{NHscaling5D}) differs from the one that is standard from the point
of view of $AdS_3/CFT_2$ duality in that we are treating the $S^1$ radius $R$ on the same
footing as the other compact coordinates, focusing on energies small compared to $1/R$.
The $AdS_3/CFT_2$  scaling limit would instead keep fixed $\tilde r / \a'$, $R$,  $V_{T^2}/\a'$ and  $V_{\tilde T^2}/\a'$.
From that point of view, the limit we have taken can be seen as an additional `very near horizon limit' as described in
\cite{Strominger:1998yg}.
\item The calculation above is valid for a black hole `mostly made up out of D1-branes'
where the D1-charge $n$ is parametrically larger than the other charges. From
(\ref{lB}) we see that the supergravity description is reliable in this regime provided that $g w  \gg 1$.
\item In the S-dual picture, where the black hole consists of wrapped fundamental strings and NS5-branes
and momentum, the relevant probe configurations are again wrapped KK monopoles, this time carrying induced
fundamental string charge by turning on momentum conjugate to $\o^{(0)}$.
\item One can also find the relevant probe configurations when other charges are large by dualizing
the relevant configurations in frame A.  For large D5-charge $w$ the probe solution is again a KK-monopole,
this time wrapped on $\tilde S^3/Z_W\times \tilde T^2$. For large momentum $N$, one finds
a D5-brane on $\tilde S^3/Z_W$ and $\tilde T^2$ with momentum along $S^1$, which can be interpreted as a giant graviton.
\end{itemize}


\section{Discussion and outlook}

In this paper, we have used U-duality and the 4D-5D connection to construct microstate probe solutions in the near-horizon
geometry of the D1-D5-P black hole. The relevant configurations are bound states of D1-branes that have expanded through the Myers effect to form a
Kaluza-Klein monopole wrapping the black hole horizon. They are static with respect to the time coordinate adapted
to the $L_0$ generator of the `left-moving' $SL(2,R)_L$, and hence are expected to correspond to bound states rather than
fragmentation modes of the system \cite{Maldacena:1998uz}.
It would interesting to study in more detail the superconformal quantum mechanics describing
the low-energy dynamics of the probes in the case $W=1$, where the right-moving $U(1)$ symmetry is enhanced to $SO(3)_R$.
In \cite{Denef:2007yt}, a refined version of the near-horizon microstate approach was proposed,  where it was argued that the probe brane
quantum mechanics arises from  the moduli space quantization of a multicentered solution. It would be of interest
to study the the analogous `deconstructed' black hole solutions and their moduli space in the case of the five-dimensional
D1-D5-P black hole. As in \cite{Denef:2007yt}, this is likely to provide a natural explanation for the fact that the probe
branes are static with respect to the specific time coordinate $\t$.

The fact that  the relevant near-horizon probes are KK-monopoles is also interesting in itself.
In the approach advocated by Lunin and Mathur \cite{Lunin:2001jy}, the nonsingular microstate geometries in the D1-D5 system
are due to the expansion of D1 and D5 branes into KK monopole supertubes \cite{Lunin:2002iz}.
It will be interesting to see if and how both approaches are related.

The probe solutions were constructed in the near-horizon  geometry of the  black hole which includes a quotient of $AdS_3$
(with the geometry of a BTZ black hole)
 and should be viewed as an averaged geometry describing an ensemble  of microscopic states, corresponding to a density
matrix in the dual CFT \cite{Maldacena:1998bw}. The probe solutions we have considered could be seen as adding
 black hole  `hair' to this averaged geometry. The  states in the ensemble we considered are characterized by microscopic
quantum numbers consisting of the wrapped KK-monopole charge and the angular momentum quantum numbers labelling
the lowest Landau level groundstates. It would be interesting to identify these states within the known ensemble
of microstates in the dual CFT. Such a comparison is obscured by the fact that the averaged near-horizon geometry
has less symmetry than the dual CFT because of the quotienting of $AdS_3$. It could therefore be interesting to study the
 limit where the momentum  $N$ is much larger than the other charges where, as one can see from (\ref{nearhorB}),
 the $AdS_3$ symmetries are approximately restored.
As we remarked in section \ref{landauB}, the relevant probe solutions in this regime are a kind of giant gravitons.
It would also be interesting to study the relation of such solutions to other giant graviton configurations constructed
recently in \cite{Mandal:2007ug,Raju:2007uj}.

\acknowledgments{I would like to thank Frederik Denef, Yuji Sugawara and Patta Yogendran for useful discussions.}

\end{document}